# Machine Learning based Context-predictive Car-to-cloud Communication Using Multi-layer Connectivity Maps for Upcoming 5G Networks

**Benjamin Sliwa[1], Robert Falkenberg[1], Thomas Liebig[2], Johannes Pillmann[1] and Christian Wietfeld[1]**
[1]Communication Networks Institute, [2]Department of Computer Science VIII
TU Dortmund University, 44227 Dortmund, Germany
e-mail: {Benjamin.Sliwa, Robert.Falkenberg, Thomas.Liebig, Johannes.Pillmann, Christian.Wietfeld}@tu-dortmund.de

***Abstract*—While cars were only considered as means of personal transportation for a long time, they are currently transcending to mobile sensor nodes that gather highly up-to-date information for crowdsensing-enabled big data services in a smart city context. Consequently, upcoming 5G communication networks will be confronted with massive increases in Machine-type Communication (MTC) and require resource-efficient transmission methods in order to optimize the overall system performance and provide interference-free coexistence with human data traffic that is using the same public cellular network. In this paper, we bring together mobility prediction and machine learning based channel quality estimation in order to improve the resource-efficiency of car-to-cloud data transfer by scheduling the transmission time of the sensor data with respect to the anticipated behavior of the communication context. In a comprehensive field evaluation campaign, we evaluate the proposed context-predictive approach in a public cellular network scenario where it is able to increase the average data rate by up to 194% while simultaneously reducing the mean uplink power consumption by up to 54%.**

## I. Introduction

Massive MTC is one of the three foundations of upcoming 5G cellular networks. In the vehicular context, cars will be exploited as mobile sensor nodes that gather data for crowdsensing-based cloud-applications like intelligent traffic forecast and distributed weather prognosis [1]. With the resulting increase of Machine-to-machine (M2M) communication, resource-efficient data transmission methods are highly demanded in order to satisfy the application demands and minimize the interference with Human-to-human (H2H) network traffic. A promising technique to address these challenges is the integration of the transmission context in the actual transmission process in order to exploit favorable channel conditions for transmitting data in a more efficient way [2] with respect to different performance indicators such as data rate, energy-efficiency and transmission duration. In earlier work, we presented the probabilistic transmissions schemes Channel-aware Transmission (CAT) and predictive CAT (pCAT) [3] that exploit knowledge about the Signal-to-noise-ratio (SNR) in order to leverage *connectivity hotspots* for achieving reliable transmissions of vehicular sensor data. In this paper, we move another step further and transit from context-awareness to context prediction. The pCAT scheme is extended analytically using a multi-metric approach allowing the joint consideration of multiple network quality indicators. Data obtained from previous measurements is exploited to construct connectivity maps (see Fig. 1) that are used as a priori information about the channel environment. With this knowledge and the use of mobility prediction, vehicles schedule their data transmissions with respect to the anticipated development of the radio channel conditions. Furthermore, the achievable data rate is predicted with machine learning methods and used itself as a transmission metric for the extended pCAT scheme. The paper is structured as follows: after discussing the related work, we present the analytical model of the extended pCAT as well as multiple mobility prediction schemes. Afterwards, the setup for the empirical evaluation is described. Finally, detailed results of the transmission schemes are presented for multiple parameterizations and application scenarios. Furthermore, the achieved accuracy for mobility prediction schemes as well as the transmission metrics is discussed. In order to achieve a high level of transparency and reproducibility, the source code of the developed measurement application as well as the raw data obtained from the experimental evaluation is provided in an Open Source way.

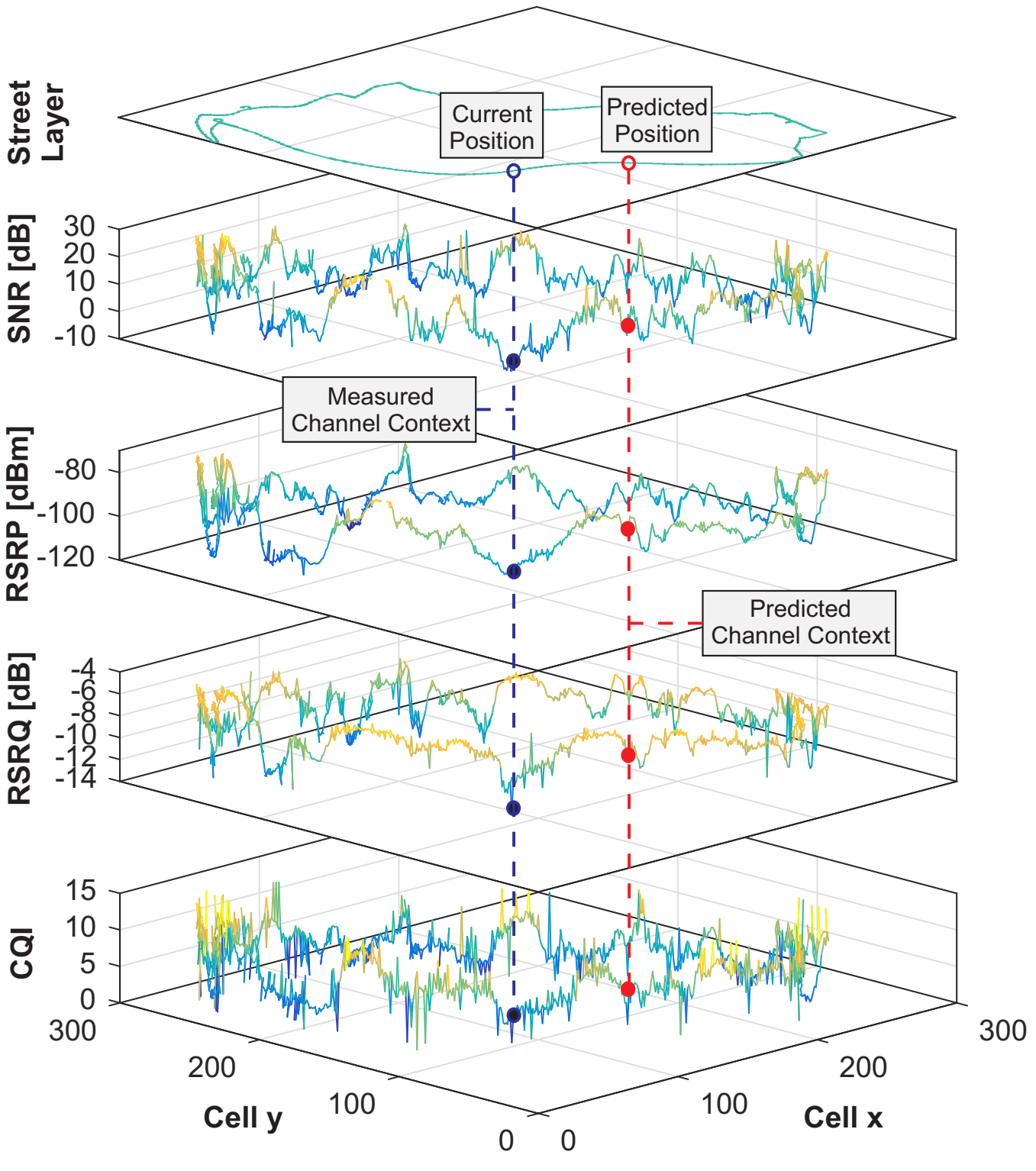

Fig. 1. Multi-layer connectivity maps as an enabler for anticipatory communication. With knowledge about the spacial dependencies of the connectivity indicators being available as a priori information, the measured channel context can be set into relation to its predicted future state.

## II. Related Work

The provision of broadband services for emerging smart city applications and Intelligent Transportation Systems (ITS) crucially depends on the real-time data exchange over a sophisticated network infrastructure. In order to ensure a reliable communication, even though the device density and traffic load continuously increase, many researchers integrate versatile approaches of data analysis into the communication systems and evolve reactive networks into intelligent and anticipating compound structures [2]. Multi-layer approaches, such as [4], provide frameworks to aggregate crowd-sensing data about road traffic, traffic lights and network conditions to guide vehicles to appropriate access networks in order to deliver services more efficiently, e.g., broadband infotainment. Other approaches utilize context information from the network for smart-parking services and road-traffic prediction to perform a context-aware dynamic vehicle routing [5].

Furthermore, the radio resource management of the Radio Access Network (RAN) requires anticipatory methods for proper handling of heterogeneous traffic types to satisfy particular needs, e.g., latency and throughput, in a large scale. Optimizing the mixture of H2H and MTC in 5G cellular networks is addressed in [6]. The main challenge is dealing with massive MTC without degrading the performance of H2H, however, still maintaining the required degree of stability for MTC. A biologically inspired approach, which models MTC and H2H as populations of predators and prey, respectively, provides a stable equilibrium for both applications. A detailed data-driven analysis of anticipatory networking strategies is given in [7]. The authors provide an optimization framework to allow network operators to save up to $50\,\%$ or resources by using either centralized or distributed prediction approaches in their base stations. In contrast to that, the proposed machine learning based extension for pCAT works decentralized on the client side and does not involve additional downlink communication for coordinating uplink transmissions. Therefore, it can be easily integrated into existing measurement applications without requiring changes in the network infrastructure.

While in theory the resulting throughput of a data transmission is the result of a deterministic process, it is a challenging task to predict the resulting value proactively in a live-system due to a wide range of hidden and dynamic variables (e.g. packet loss, channel stability, transport-layer interdependencies and available spectral resources), especially in the presence of mobility. In the recent years, great advancements have been achieved by bringing together engineering expertise for data acquisition and identification of relevant features and machine learning based data analysis [8]. In [9] the authors utilize machine learning for high-accuracy data rate predictions, based on passive connectivity indicators in conjunction with the estimated activity of other participants in the same cell. The data is obtained by blind-decoding the Long Term Evolution (LTE) Physical Downlink Control Channel (PDCCH) in real time, which provides the User Equipment (UE) a global overview of the current cell load and the momentary degree of competition for radio resources. Other approaches exploit connectivity maps for an optimized utilization of spectral resources, especially for vehicular applications [10]. Such maps can be obtained from sparse sampling the environment and performing a kernel-based reconstruction of coverage maps [11], which can be downloaded to the mobile device for client-based decision-making.

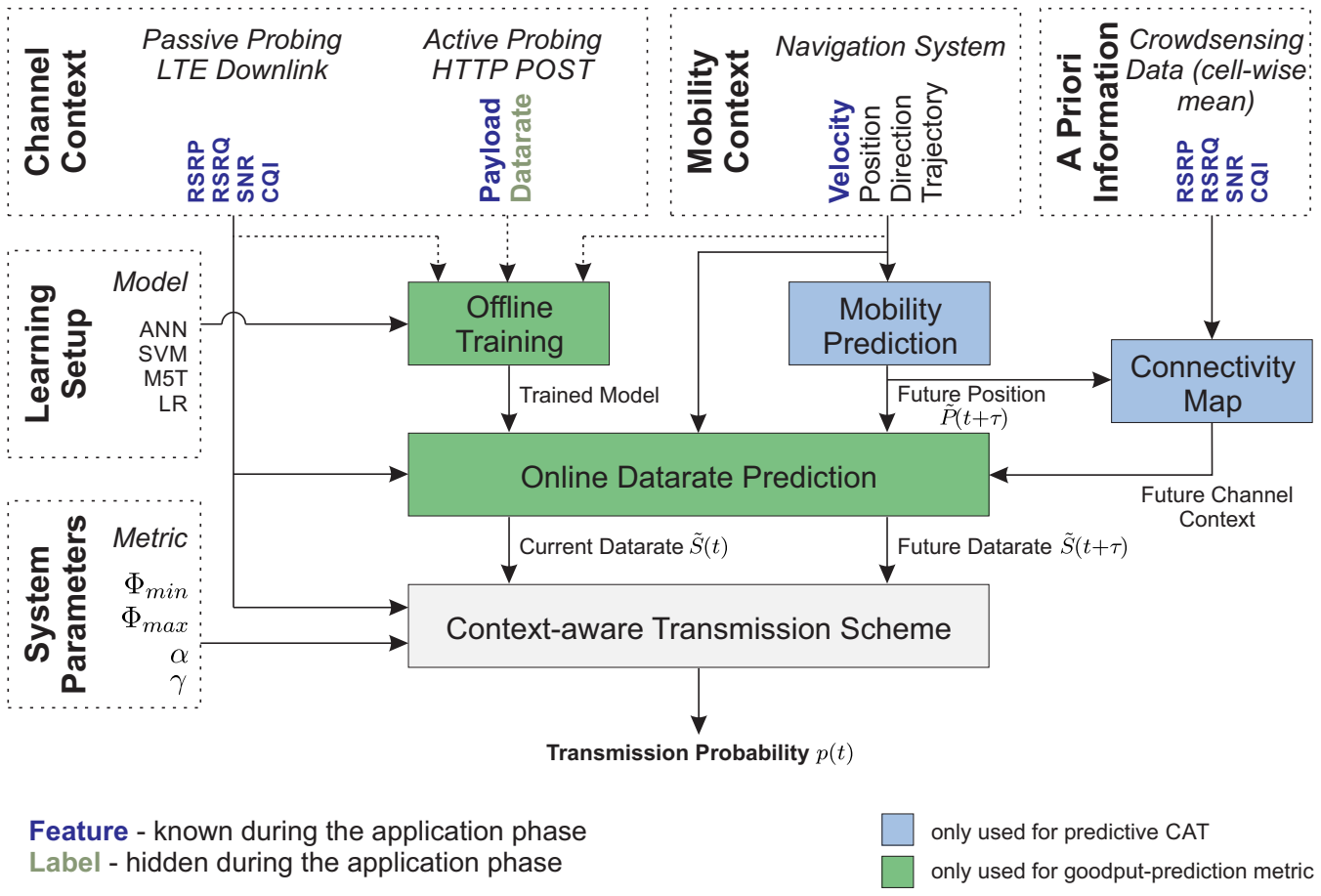

Fig. 2. Architecture model of the proposed transmission scheme. Different context layers are brought together with a priori information about the radio channel obtained from previous measurements. The dashed connections are only required during the training phase of the selected machine learning model.

## III. Solution Approach

The overall architecture model of the proposal that operates on the application layer is illustrated in Fig. 2. Obtained sensor data is buffered locally until a transmission decision is made for the whole buffer. The decision itself is based upon currently measured information and predicted data obtained from the connectivity map and can either use a single context parameter or machine learning based data rate prediction as a metric. Fig. 3 provides an example that compares naive periodic data transfer to the proposed transmission scheme.

In the first step, a mechanism to predict the achievable data rate $\tilde{S}(t)$ based on the measured channel context parameters is derived by training different machine learning models (Artificial Neural Network (ANN), Support Vector Machine (SVM), M5 Decision Tree (M5T) and Linear Regression (LR)) (see Sec. III-1). The channel context is formed by the reported values of the passive LTE downlink indicators Reference Signal Received Power (RSRP), Reference Signal Received Quality (RSRQ), SNR, Channel Quality Indicator (CQI) as well as the payload size of the data packet, which represent the features of the machine learning process. During the training phase, the achieved data rate of active Hypertext Transfer Protocol (HTTP) POST transmissions is monitored and used as the prediction label.

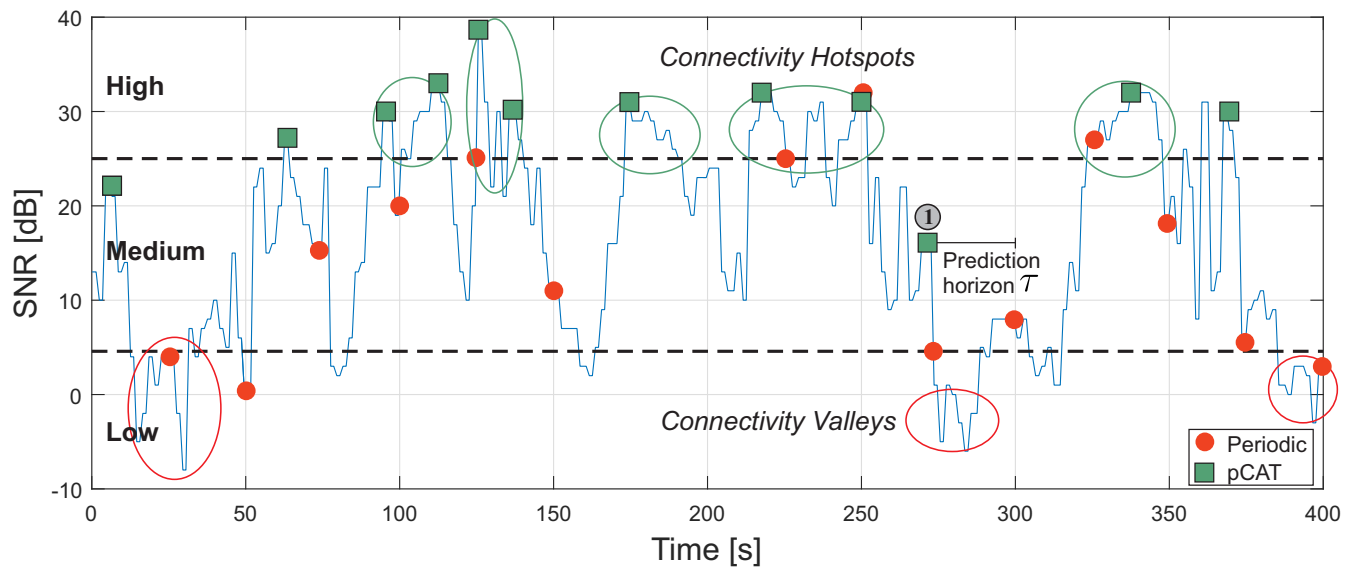

Fig. 3. Example temporal behavior of the different transmission schemes. While naive periodic transmissions are executed regardless of the channel quality, the proposed pCAT determines a transmission probability that takes the current and predicted channel quality into account. At (1), pCAT transmits data at medium network quality as it expects the vehicle to pass a connectivity valley in the near future.

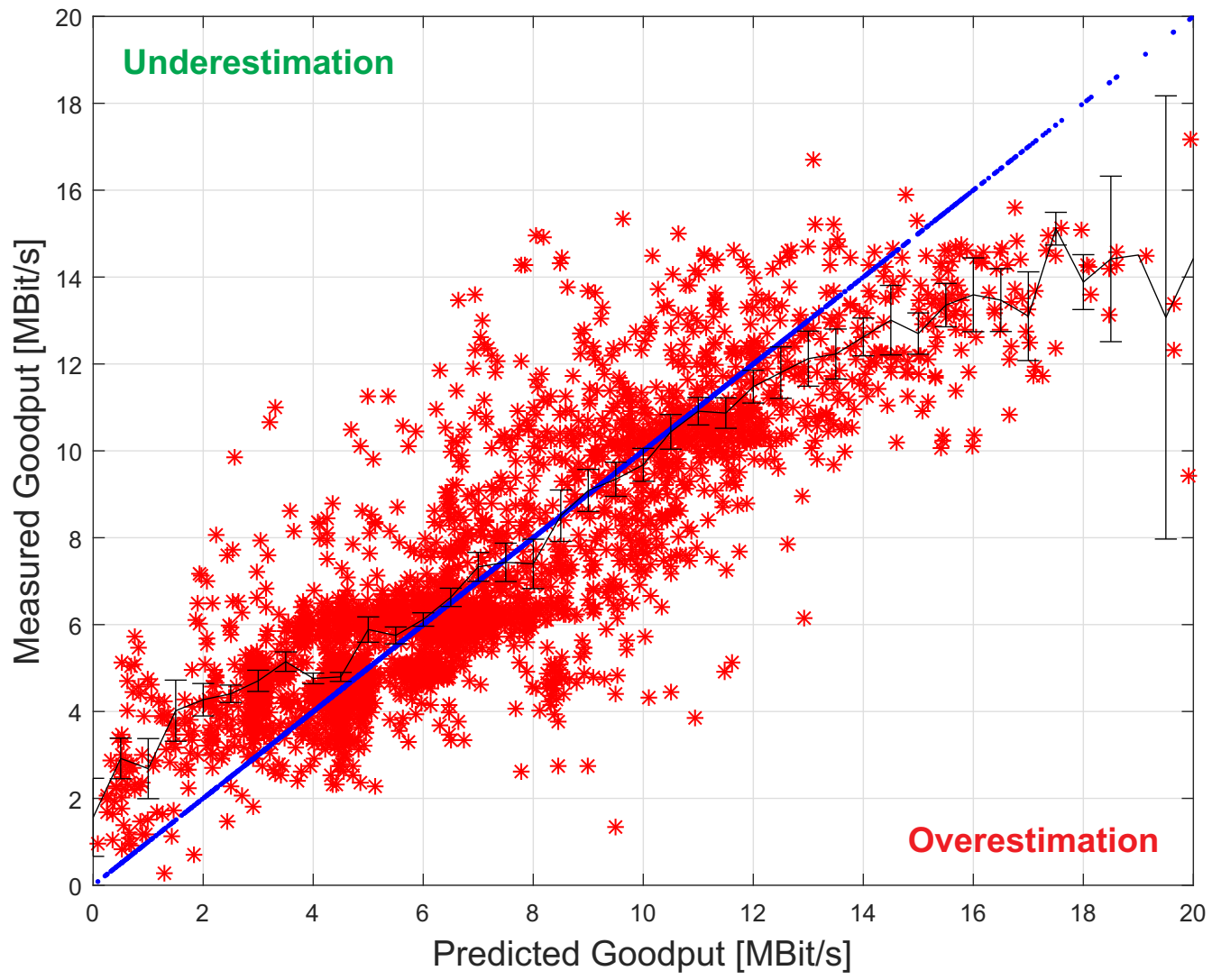

Fig. 4. Accuracy of the data rate prediction with the M5 decision tree based on [12]. The blue dots illustrate the behavior of an optimal prediction as a reference. The black curve shows the 0.95 confidence-interval of the mean value.

For obtaining knowledge about the future channel context from the connectivity map, vehicles need to first determine their future position, which is then used to find the matching cell entry of the map that contains the aggregated context information. For this purpose, mobility prediction (cf. Sec. III-A) is applied to estimate the future position $\vec{P}(t+\tau)$ for a defined *prediction horizon* $\tau$, based on the current mobility parameters and optional trajectory knowledge.
The online data rate prediction is then performed for the measured channel quality parameters as well as for the predicted channel context parameters. For this purpose, the machine learning model with the highest achieved prediction accuracy is applied. With the availability of estimations about the currently achievable data rate $\tilde{S}(t)$ and the future data rate $\tilde{S}(t+\tau)$, the context-aware transmission (cf. Sec. III-B) is then performed based on a defined metric $\Phi$, which computes a transmission probability for the current time $t$. The process is periodically evaluated with a fixed update interval $t_p$.
If no valid location information is available or if one of the prediction steps fails, the pure probabilistic basic CAT scheme is used as a fallback mechanism that is only considering the currently measurable context information. In the following subsections, the individual components and their relevant parameters are described in details.

*1) Machine Learning based Data Rate Prediction:* In [12], we have compared the accuracy of the data rate prediction for multiple machine learning models (ANN, SVM, M5T, LR). In the following, only the best performing model (M5T) is applied in the further evaluations. Fig. 4 shows the behavior of the prediction accuracy with Mean Absolute Error (MAE) = 1.33 MBit/s and Root Mean Square Error (RMSE) = 1.81 MBit/s. The blue curve represents a perfect prediction, where the predicted values are equal to the achieved measured values. While there is a two-sided error range, only the *overestimation* area (lower triangle of the plot) is considered harmful for the data transmission as *underestimations* (upper triangle of the plot) even result in a higher gain in the data rate than expected. The great advantage of using machine learning for channel quality assessment is that it does not only consider the explicitly defined features RSRP, RSRQ, SNR, CQI, payload size and velocity but also implicitly integrates *hidden parameters* into the evaluation process. With considering the payload size, the time-stability of the communication channel and cross-layer dependencies such as the slow start of the Transmission Control Protocol (TCP) and the payload-overhead-ratio have an impact on the estimated data rate. Similarly, the RSRQ value contains hidden information about the available cell resources as well as the interference situation.

### A. Mobility Prediction

For the estimation of the future channel context, knowledge about the future location is a basic requirement. Therefore, different approaches for predicting the future vehicle position $\vec{P}(t+\tau)$ for a defined prediction horizon $\tau$ are applied which differ in the utilized sensor information. For simplicity, the following calculations are presented in a cartesian coordinate system. For using raw Global Positioning System (GPS) coordinates, all calculations have to be performed in the orthodromic dimension [13].

*1) GPS-based Extrapolation:* The simplest prediction approach utilizes the available information from the GPS module of the UE and uses the current values for position $\vec{P}(t)$, angular direction $\lambda$ and the velocity $v$ to extrapolate the future location with Eq. 1

$$\vec{P}(t+\tau) = \vec{P}(t) + \begin{pmatrix} \sin(\frac{\Pi}{2}) \cdot \cos\left(\frac{\lambda \cdot \Pi}{180}\right) \\ \sin(\frac{\Pi}{2}) \cdot \sin\left(\frac{\lambda \cdot \Pi}{180}\right) \end{pmatrix} \cdot \tau \cdot v \qquad (1)$$

Since the prediction relies of the assumption of a static direction for the duration of $\tau$, it is severely impacted by the turns the vehicle performs on its route, especially in urban environments with many road junctions.

*2) Leveraging Trajectory-knowledge from the Navigation System:* Precise predictions of the vehicle's future location can be achieved by integrating knowledge about the planned trajectory into the prediction process. Even if this information is not explicitly available, it may be derived exploiting the regularities of human behavior [14], since most human trajectories can be predicted with a high accuracy as people usually follow daily routines and visit the same places regularly. A single *trip* is described by its starting point and destination (e.g. the way from home to work). It is assumed that multiple traces for the same trip are already available and that the network quality has been monitored continuously and was used to build the connectivity map along the track (cf. Sec. IV-A). The mean trajectory over all traces for the same trip is then computed based on the algorithm proposed in [15]. With this information, the predicted position is computed with an iterative process. At first, the distance potential $\tilde{D} = v \cdot \tau$ is determined. Afterwards, the distance $d_{i,j}$ between consecutive waypoints $\vec{W}_i$ and $\vec{W}_{j=i+1}$ on the trajectory $T$ is computed and the travelled distance $D$ is incremented by $d_{i,j}$. Once $D$ exceeds $\tilde{D}$, the final position is computed by interpolation with Eq. 2.

$$\vec{P}(t+\tau) = \vec{W}_i + \frac{\vec{W}_j - \vec{W}_i}{||\vec{W}_j - \vec{W}_i||} \cdot \left(D - \tilde{D} - d_{i,j}\right) \qquad (2)$$

For the lookup of the context entry from the connectivity map, the position is converted to a map index $m = \lfloor \frac{\vec{P}(t+\tau)}{c} \rfloor$ with a defined cell size $c$ that describes the aggregation area.

*3) Prediction based on a Reference Trace:* A simpler approach is to declare a single trip as a reference and then apply the same prediction mechanism as in Sec. III-A2. In contrast to the other approaches, the measured values are not grouped in cell-structures but mapped to discrete points (with respect to the context sampling frequency $f_{context}$) on the geographical track. The channel quality indicators are obtained from the reference measurement and not from the connectivity map. While this approach is easy to implement, it does not perform any cell-wise data aggregation. In Sec. V-A the different approaches are compared with respect to their achievable prediction accuracy for the individual indicators.

### B. Probabilistic Context-predictive Transmission of Vehicular Sensor Data

Once the future vehicle location is known and the predicted context information has been obtained from the connectivity map, the probabilistic transmission scheme is applied. In order to allow the combination of multiple metrics and provide a mechanism for comparing metrics that have different value ranges, a generic metric $\Phi$ is introduced, which is defined by a maximum value $\Phi_{max}$ and a minimum value $\Phi_{min}$. Furthermore, the metric exponent $\alpha$ controls how much the resulting transmission probability should be depending on high metric values.

In the first step, the measured metric value $\Phi(t)$ is transformed into the normed current metric value $\Theta(t)$ with Eq.3.

$$\Theta(t) = \frac{\Phi(t) - \Phi_{min}}{\Phi_{max} - \Phi_{min}} \quad (3)$$

With $\tilde{\Phi}(t + \tau)$ describing the future metric value obtained from the connectivity map, the anticipated gain $\Delta\Phi(t)$ is then computed using Eq. 4.

$$\Delta\Phi(t) = \tilde{\Phi}(t + \tau) - \Phi(t) \quad (4)$$

Finally, the transmission probability $p_\Phi(t)$ is computed with respect to the defined timeout values $t_{min}$ and $t_{max}$ that guarantee a minimum required packet size and a maximum allowed buffering delay with Eq. 5. The pCAT-specific exponent $z$ (c.f. Eq. 6) is used to increase the transmission probability if the channel quality decreases and to delay transmissions if the channel quality improves.

$$p_\Phi(t) = \begin{cases} 0 : \Delta t \leq t_{min} \\ \Theta(t)^{\alpha \cdot z} : t_{min} < \Delta t < t_{max} \\ 1 : \Delta t > t_{max} \end{cases} \quad (5)$$

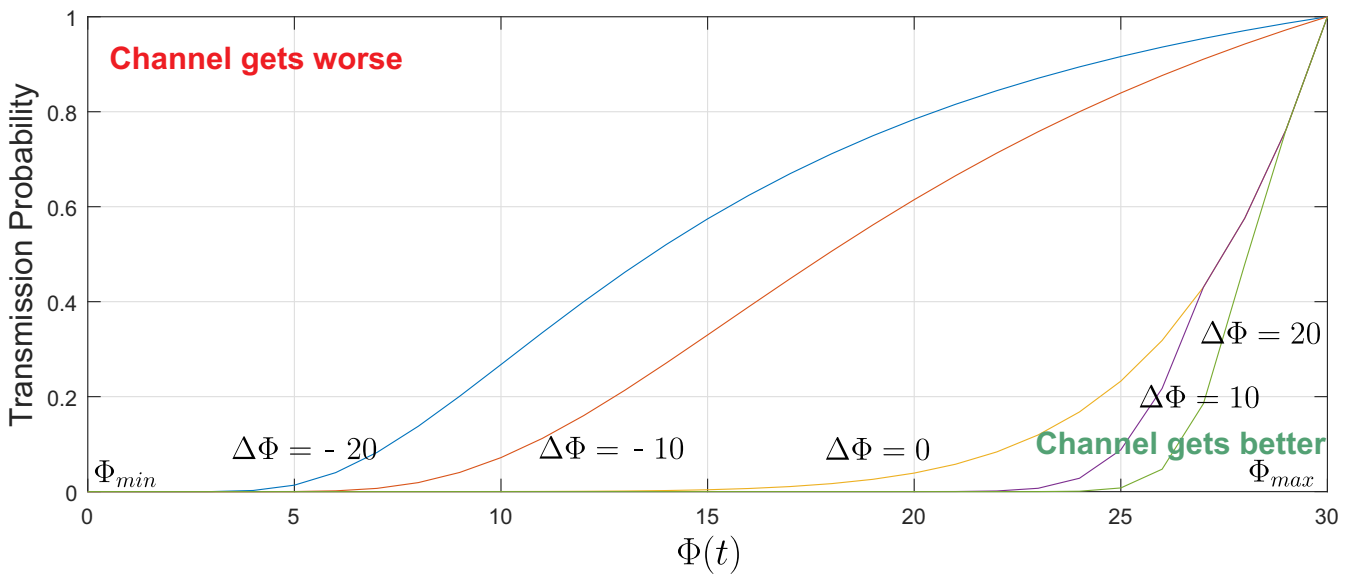

Fig. 5. Analytical behavior of the probabilistic pCAT transmission scheme. Transmissions are performed with a higher probability if the channel quality decreases ($\Delta\Phi \leq 0$) and are delayed if the anticipated channel quality is high ($\Delta\Phi > 0$).

$$z = \begin{cases} \max\left(\left|\Delta\Phi(t) \cdot (1 - \Theta(t)) \cdot \gamma\right|, 1\right) : \Delta\Phi > 0 \\ \left(\max\left(\left|\Delta\Phi(t) \cdot \Theta(t) \cdot \gamma\right|, 1\right)\right)^{-1} : \Delta\Phi \leq 0 \end{cases} \quad (6)$$

If the connectivity map does not contain an entry for the estimated position $\vec{P}(t + \tau)$, the non-predictive CAT scheme as described in [12] is applied as a fallback mechanism. Fig. 5 shows the behavior of the analytical pCAT model for different values of $\Delta\Phi$. For $\Delta\Phi = 0$, the transmission schemes behaves equally to simple non-predictive CAT.

## IV. Setup of the Empirical Evaluation

In this section, the setup for the empirical evaluation campaign and the methodology of the data analysis are presented.

The default parameters for the evaluation scenario are defined in Tab. I and the parametrization of the considered transmission metrics is shown in Tab. II. A virtual sensor application generates 50 kByte of data per second and the transmission decision for the whole local data buffer is performed second-wise.

TABLE I
PARAMETERS OF THE REFERENCE SCENARIO

| **Parameter** | **Value** |
|---|---|
| Sensor frequency $f_{sensor}$ | 1 Hz |
| Sensor payload size $s_{sensor}$ | 50 kByte |
| Transmission probability update interval $t_p$ | 1 s |
| $t_{min}$ | 10 s |
| $t_{max}$ | 120 s |
| Prediction horizon $\tau$ | $\{10, 30, 60\}$ s |
| Connectivity map cell size $c$ | 25 m$^2$ |
| Context sampling frequency $f_{context}$ | 1 Hz |

TABLE II
PARAMETERS FOR THE CONSIDERED METRICS

| | $\Phi_{RSRP}$ | $\Phi_{RSRQ}$ | $\Phi_{SNR}$ | $\Phi_{CQI}$ | $\Phi_{M5T}$ |
|---|---|---|---|---|---|
| min | -120 dBm | -11 dB | 0 dB | 2 | 0 MBit/s |
| max | -70 dBm | -4 dB | 30 dB | 16 | $\{15, 18\}$ MBit/s |
| $\alpha$ | 8 | 8 | 8 | 8 | 8 |
| $\gamma$ | 0.3 | 2.14 | 0.5 | 1.07 | 1 |

### A. Evaluation Scenario

The map of the real-world scenario used for the performance evaluation is shown in Fig. 6, consisting of a highway track and a suburban track. For each parametrization of the transmission schemes, the performance was evaluated 5 times on each of the tracks. Overall, more than 6000 HTTP POST transmissions were performed within a total driven distance of more than 2000 km. The actual transmissions were performed with a Samsung Galaxy S5 Neo (Model SM-G903F) smartphone in a public cellular network with a developed Android-based measurement application that is available in an Open Source manner[1]. Furthermore, the raw data obtained from the measurements is provided at [16].

The channel quality data to build the connectivity map and the mobility data for the trajectory-based prediction is based on measurements from 90 previous drive tests that were performed in the same scenario in [12].

[1] Available at https://github.com/BenSliwa/MTCApp

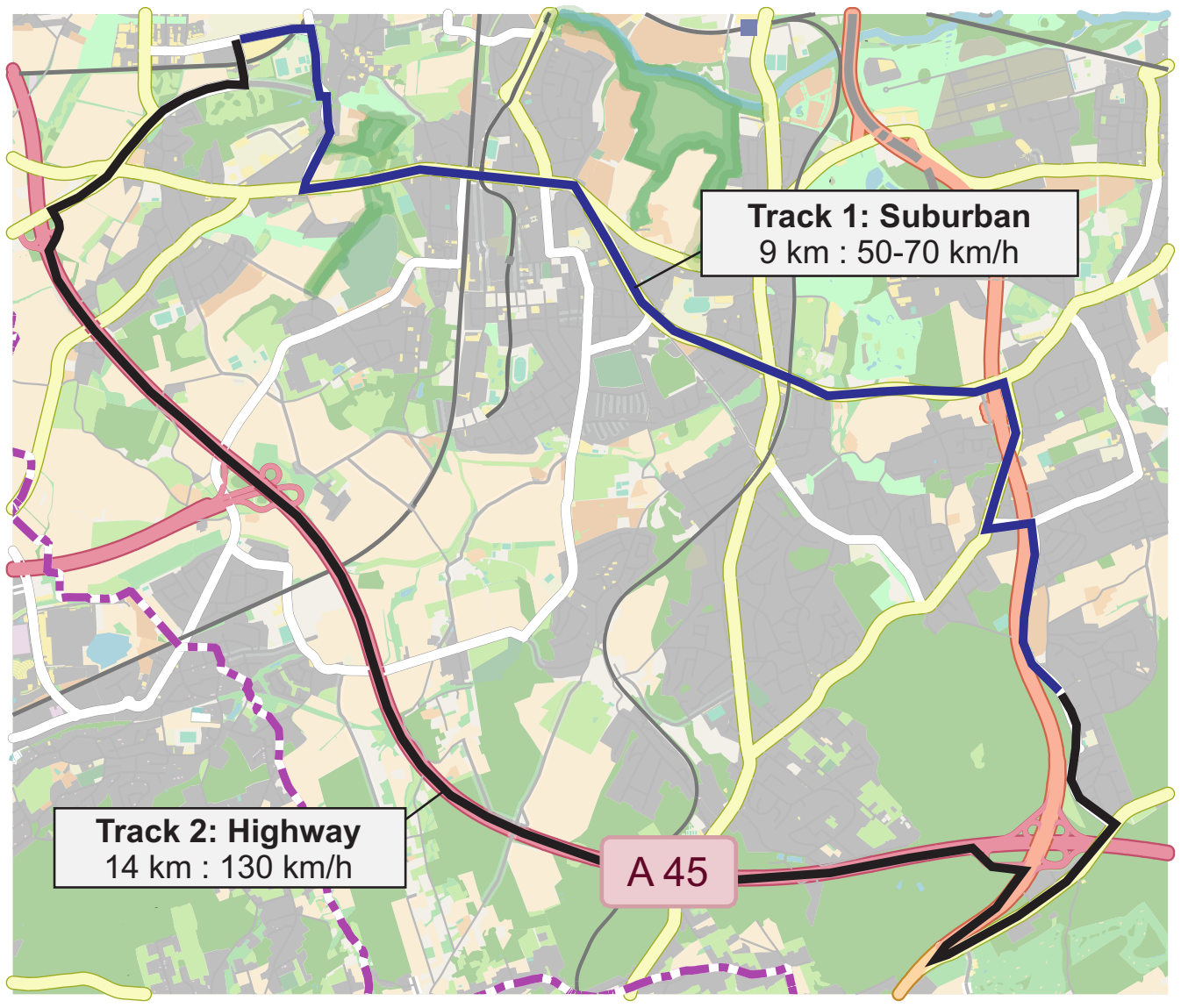

Fig. 6. Overview of the 23km long test route containing different street types and speed limitations. (Map: ©OpenStreetMap contributors, CC BY-SA.). *Source:* Adapted from [12].

The values for the weighting factor $\gamma$ have to be chosen with respect to the metric's value range and its granularity. For allowing a fair comparison among the different metrics, $\gamma$ has been chosen with respect to the definition of $\Phi_{SNR}$ metric and it's respective value range with Eq. 7.

$$\Phi_{i,\gamma} = \Phi_{SNR,\gamma} \cdot \frac{\Phi_{SNR,max} - \Phi_{SNR,min}}{\Phi_{i,max} - \Phi_{i,min}} \quad (7)$$

### *B. Analysis of the Uplink Power Consumption with the Context-aware Power Consumption Model (CoPoMo)*

The accurate determination of the isolated power consumption caused by the uplink data transmissions is a challenging task and is most likely performed in a laboratory environment and not within a mobile setup. However, for estimating the uplink power consumption in the highly-mobile scenario, the CoPoMo can be applied that has been proposed and validated in [17]. The model takes the different states of the power amplifier into account and maps the transmission power to discrete power states. Since those states are highly specific for the actual used device, the device characteristics need to be obtained using a laboratory measurement setup. Since the measurement device does not report the actual transmission power value, a machine learning based prediction method is applied [18]. Fig. 7 shows the device characteristics for the measurement device at 800 MHz and 2600 MHz. It can be seen that both curves can be approximated by two linear functions that are separated by the parameter $\gamma$. In the following, the model is applied to estimate the overall power consumption for the performed uplink transmissions.

## V. Results of the Experimental Performance Evaluation

In this section, the results of experimental performance evaluation are presented. At first, the prediction accuracy of relevant performance indicators is evaluated. Afterwards, a performance comparison is provided for the considered context-predictive transmissions schemes that use either a single indicator or the machine learning based data rate prediction as a metric.

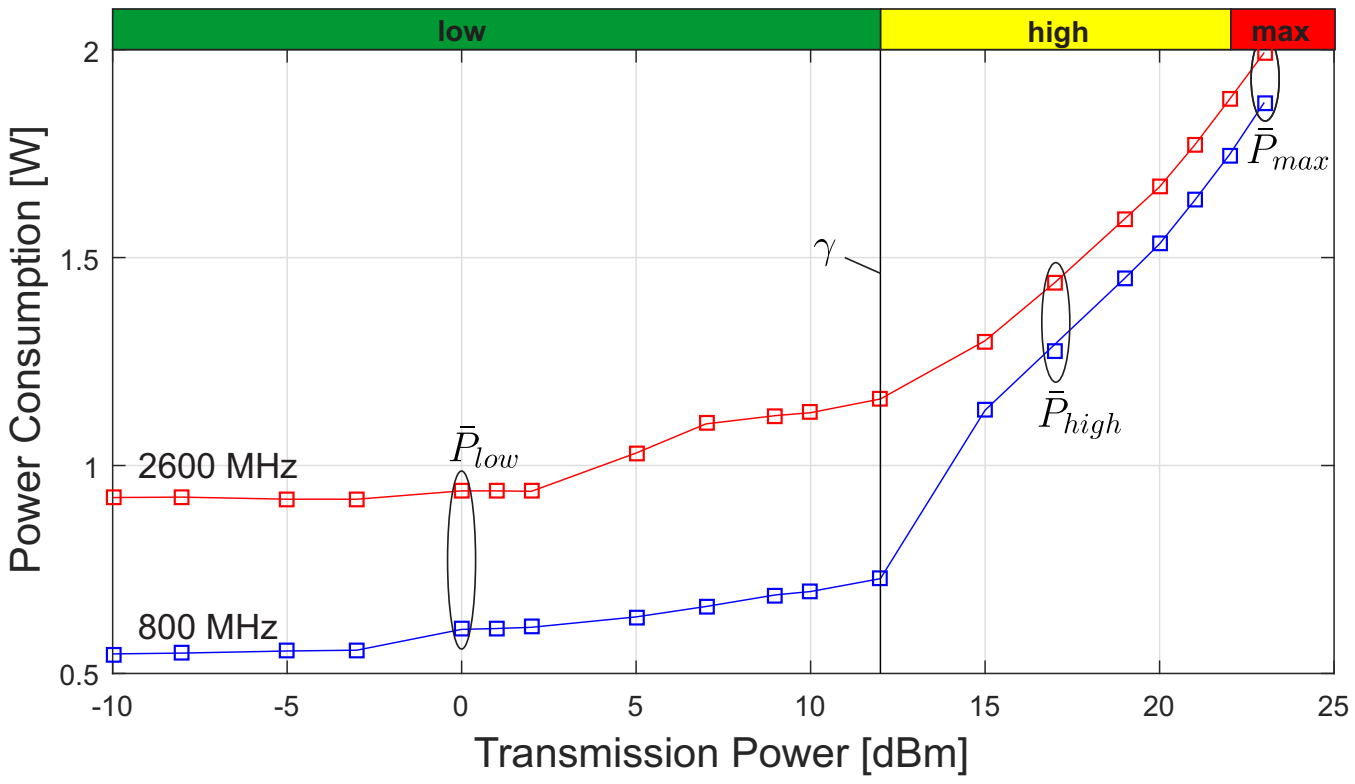

Fig. 7. CoPoMo device characteristics for the measurement device obtained for laboratory measurements. The colored boxes above the plot illustrate the CoPoMo states with respect to the transmission power.

### *A. Evaluation of the Prediction Accuracy using Connectivity Maps*

Since the transmission schemes base their decisions on the predicted context information, the ability to obtain accurate predictions is crucial for the overall system performance. Multiple errors are system-immanent and should be minimized:

- Positioning inaccuracy of GPS, especially in the presence of obstacles and urban canyons.
- Non-static influences like traffic signals and obstructions caused by other traffic participants are not considered by the mobility prediction.
- Aggregation loss due to cell-wise aggregation of measured data within the connectivity maps, especially for indicators that are highly influenced by dynamic effects like multipath-fading and interference.

Fig. 8 shows the achieved prediction accuracy for the considered indicators with different prediction schemes and prediction horizons. It is obvious that the integration of trajectory information from the navigation system reduces the magnitude of the position prediction error significantly. The pure GPS-based extrapolation suffers severely from direction changes of the vehicle. For the speed-dependency of the prediction error, three different regions can be identified. Up to about 70 km/h (urban/suburban road traffic (1)), the error is proportional to the velocity. For higher velocities, the road characteristics change due to the highway scenario (2) and the direction of road segments is constant for longer time periods, resulting in a decrease phase of the error. However, for velocities above 90 km/h (3), the higher prediction distance becomes the dominant error source again. The achievable accuracy has to be regarded in relation with the prediction success ratio since the forecast is useless for the actual transmission decision if the connectivity map does not contain context information for the falsely predicted cell. Therefore, in the following considerations, the GPS-based prediction is not evaluated anymore, as its prediction failure ratio is unacceptable high. While the RSRP and the RSRQ only show slight differences between the trajectory-based and the reference-track-based prediction, this does not apply to the SNR and the CQI that have a high dependency to the dynamics of the channel environment. Consequently, both values achieve an *aggregation gain* through the cell-wise averaging by the connectivity maps.

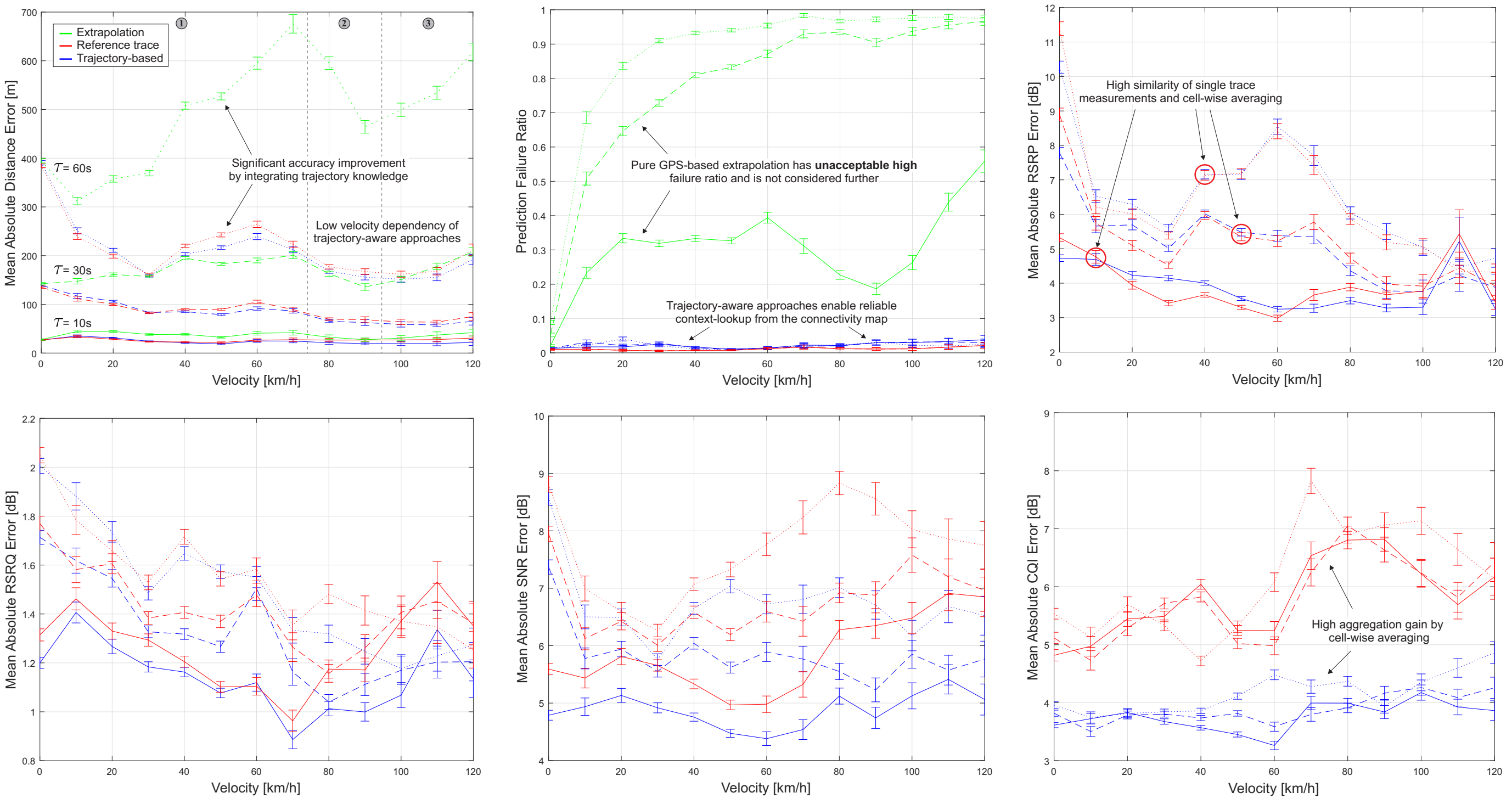

Fig. 8. Prediction accuracy for the vehicle's mobility and the passive downlink indicators of the LTE signal for the considered prediction methods with different predictions horizons in relationship to the vehicle's velocity. All curves show the 0.95 confidence interval of the mean value.

### *B. Performance Evaluation of the Proposed Transmission Scheme with Different Metrics*

Fig. 9 provides a comparison of the resulting data rate, age of information and uplink power consumption for different variants of the proposed pCAT that use a single downlink indicator as the transmission metric. The highest data rate and energy-efficiency gains are achieved for the SNR- and the RSRP-based approaches. While the regular CAT has to pay for the increased data rate with an increased age of information, the trade-off between data rate and introduced delay is significantly better for the predictive scheme. Nevertheless, the results have to be set into relation to the actual application requirements. While crowdsensing-based services rely on up-to-date information for analysis and prognosis, often, real-time capabilities are not required and the data age is allowed to be in the range of several minutes. Fig. 10 provides a performance comparison for the machine learning based data rate prediction metric $\Phi_{M5T}$ with different values for $\Phi_{max}$ and $\tau$. For $\Phi_{max} = 18$ MBit/s and $t = 30$ s, the highest performance increase is achieved with data rate gain of 194% and reduction of the mean uplink power consumption by 54%. Interestingly, the pCAT approach is able to achieve high throughput values even for higher values of $\tau$, where the context prediction is significantly distorted by uncertainties of the mobility prediction (cf. Fig. 8). While the computation of the transmission probability only evaluates the context for two discrete points of time $t$ and $t + \tau$, the computation itself is performed second-wise with $t_p$. Therefore, it behaves similar to a moving window with the car itself moving forward on the trajectory. Consequently, the transmission scheme is much more influenced by the proability of the car experiencing the predicted context within any point of time in the remaining time until $t_{max}$ is reached than to have a perfect prediction for the actual time point.

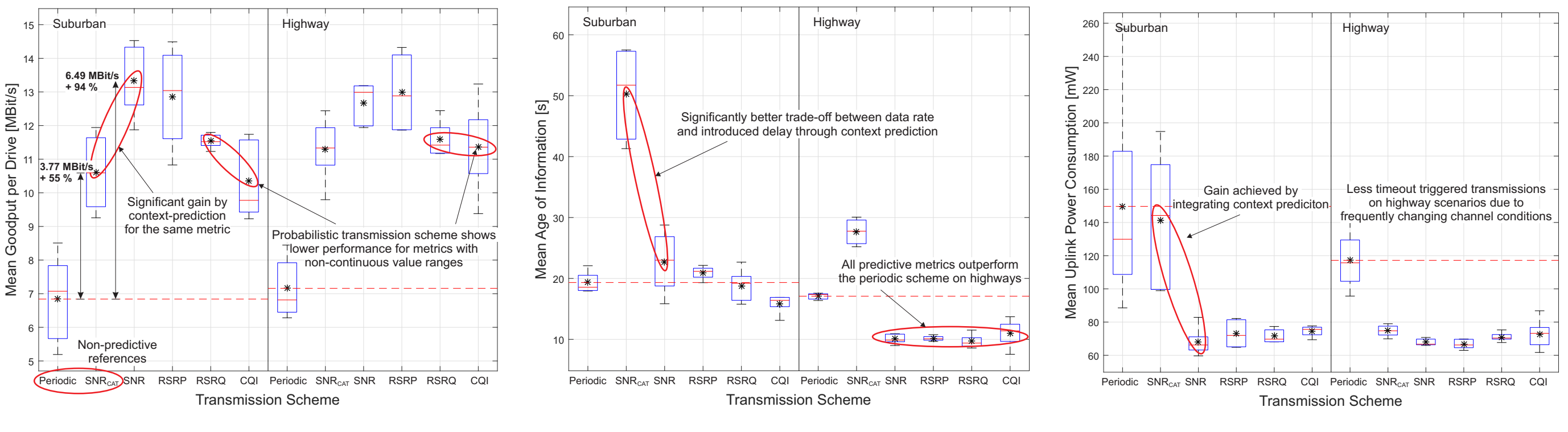

Fig. 9. Performance comparison of the proposed transmission scheme with different single indicator metrics for a fixed prediction horizon $\tau = 10$ s. The results for periodic transmission and non-context-predictive SNR-based CAT are shown as reference.

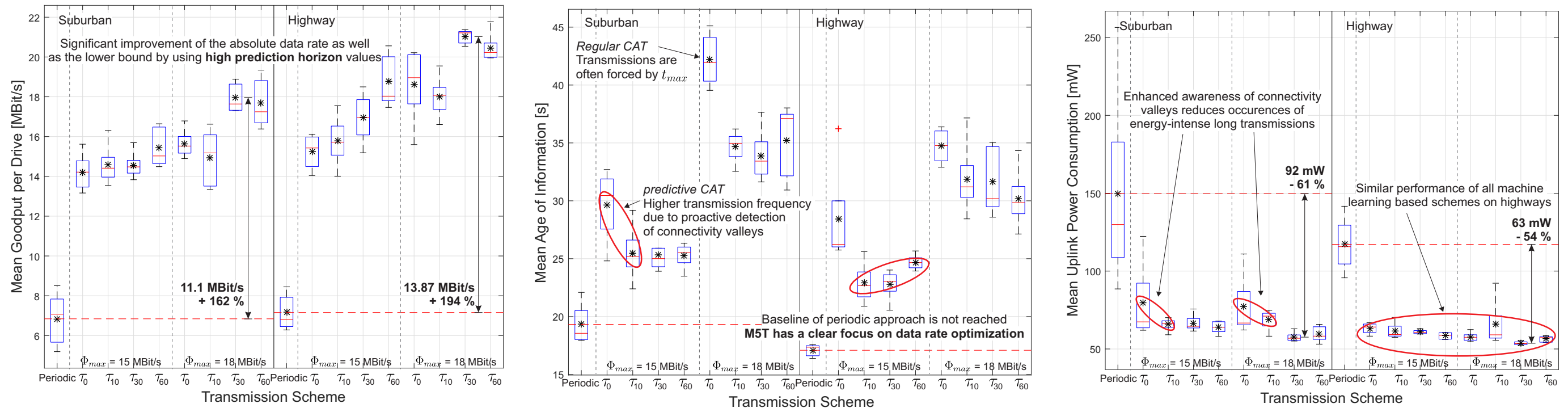

Fig. 10. Performance comparison of the proposed transmission scheme with machine learning based data rate prediction using the $\Phi_{M5T}$ metric. The results for periodic transmission and non-context-predictive M5T-based CAT ($\tau$ = 0 s) are shown as reference.

Since the machine learning based approach integrates the payload size of the data packet into the prediction process, it implicitly considers effects that have a strong dependency to the packet size and the resulting transmission duration into the actual data transmission decision. In particular, the payload-overhead-ratio and the slow start of TCP benefit from larger packets, whereas the time stability of the channel can be elaborated better for shorter transmission durations.

## VI. Conclusion

In this paper, we presented a machine learning enabled approach for context-predictive transmission of vehicular sensor data that relies on mobility prediction and channel quality anticipation. Since the properties of the radio communication channel have a severe impact on the resource-efficiency of the data transfer, naive periodic data transmissions suffer from packet loss which results in low data rate and high power consumption. Within our comprehensive empirical evaluations in public cellar networks, we were able to show that context-predictive communication is able to significantly increase the end-to-end goodput while simultaneously saving energy due to exploitation of connectivity hotspots and avoidance of retransmissions. While there is a trade-off between the achievable gains and the introduced additional buffering delay, the abstract metric definition allows to configure the proposed transmission scheme for prioritising one of the aspects with respect to the application requirements. The raw measurement data as well as the developed measurement application is provided in an Open Source way in order to increase the transparency and reusability of the performed evaluation campaign. In future work, we will further optimize the transmission scheme by optimizing the prediction accuracy for the mobility behavior as well as the channel quality estimation.

## Acknowledgment

Part of the work on this paper has been supported by Deutsche Forschungsgemeinschaft (DFG) within the Collaborative Research Center SFB 876 “Providing Information by Resource-Constrained Analysis”, projects A4 and B4. Thomas Liebig received funding from the European Union Horizon 2020 Programme (Horizon2020/2014-2020), under grant agreement number 688380 “VaVeL: Variety, Veracity, VaLue: Handling the Multiplicity of Urban Sensors”.